 \documentclass[amssymb, reprint, showpacs, aip]{revtex4-1}				  % use default format for two column to emulate published

\usepackage{graphicx}			% Include figure files
\usepackage{dcolumn}			% Align table columns on decimal point
\usepackage{bm}					% bold math
\usepackage[colorlinks=true,linkcolor=blue,citecolor=blue]{hyperref}
\usepackage{float}			% Include float files         https://www.ctan.org/pkg/float?lang=en
\usepackage{amsmath,amsfonts}	% popular packages from the American Mathematical Society
\usepackage{url}					% https://www.ctan.org/pkg/url
\usepackage{setspace}		% Sets spacing    https://www.ctan.org/pkg/setspace
\usepackage{lineno}   			% get from ctan      https://www.ctan.org/pkg/lineno
\usepackage{subfigure}
\usepackage{multirow}

\begin{document}

\begin{space}      	% {2.0}

%\preprint{JASA/123}		%  if you want want this message to appear in upper left corner of title page

\title{Source localization in an ocean waveguide using supervised machine learning}      % don't need 2 lines, example to show linebreak \\

\author{Haiqiang Niu}			
%%\email{hniu@ucsd.edu}
\thanks{Also at State Key Laboratory of Acoustics, Institute of Acoustics, Chinese Academy of Sciences, Beijing 100190, People's Republic of China. Electronic mail:  hniu@ucsd.edu}
\author{Emma Reeves}
\author{Peter Gerstoft}			
\affiliation{Scripps Institution of Oceanography, University of California San Diego, La Jolla, California 92093-0238, USA}

\date{\today}

\begin{abstract}
Source localization in ocean acoustics is posed as a machine learning problem in which data-driven methods learn source ranges directly from observed acoustic data. The pressure received by a vertical linear array is preprocessed by constructing a normalized sample covariance matrix (SCM) and used as the input for three machine learning methods: feed-forward neural networks (FNN), support vector machines (SVM) and random forests (RF). The range estimation problem is solved both as a classification problem and as a regression problem by these three machine learning algorithms. The results of range estimation for the Noise09 experiment are compared for FNN, SVM, RF and conventional matched-field processing and demonstrate the potential of machine learning for underwater source localization.
\end{abstract}

\pacs{43.60.Np, 43.60.Jn, 43.30.Wi}		% From   http://scitation.aip.org/upload/ASA/JASA/JASAAE.pdf  OPTIONAL

%Running title: Source localization using machine learning

\maketitle

%  End of title page -------------------------------------------------------------------------------------------------------------------------------- %

%\linenumbers				% start linenumbers here

\section{\label{sec:1} Introduction}

Machine learning is a promising method for locating ocean sources because of its ability to learn features from data, without requiring sound propagation modeling. It  can be used for unknown environments.

Acoustic source localization in ocean waveguides is often solved with matched-field processing (MFP).\cite{Jensen}${}^{-}$\cite{Dosso3} Despite the success of MFP, it is limited in some practical applications due to its sensitivity to the mismatch between model-generated replica fields and measurements. MFP gives reasonable predictions only if the ocean environment can be accurately modeled. Unfortunately, this is difficult because the realistic ocean environment is complicated and unstable.

An alternative approach to the source localization problem is to find features directly from data.\cite{Fialkowski}${}^{-}$\cite{Song} Interest in machine learning techniques has been revived thanks to increased computational resources as well as their ability to learn nonlinear relationships. A notable recent example in ocean acoustics is the  application of nonlinear regression to source localization.\cite{Lefort} Other machine learning methods have obtained remarkable results when applied to areas such as speech recognition,\cite{Hinton} image processing,\cite{Krizhevsky} natural language processing,\cite{Collobert} and seismology.\cite{Sharma}${}^{-}$\cite{Riahi} Most underwater acoustics research in machine learning is based on 1990s neural networks. Previous research has applied neutral networks to determine the source location in a homogeneous medium,\cite{Steinberg} simulated range and depth discrimination using artificial neural networks in matched-field processing,\cite{Ozard} estimated ocean-sediment properties using radial basis functions in regression and neural networks,\cite{Caiti1}${}^{,}$\cite{Caiti2} applied artificial neural networks to estimation of geoacoustic model parameters,\cite{Stephan}${}^{,}$\cite{Benson}  classification of seafloor\cite{Michalopoulou2} and whale sounds.\cite{Thode} 

This paper explores the use of current machine learning methods for source range localization. The feed-forward neural network (FNN), support vector machine (SVM) and random forest (RF) methods are investigated. There are several main differences between our work and previous studies of source localization and inversion:\cite{Lefort}${}^{,}$\cite{Steinberg}${}^{-}$\cite{Thode}

1. Acoustic observations are used to train the machine learning models instead of using model-generated fields.\cite{Steinberg}${}^{,}$\cite{Ozard}${}^{,}$\cite{Caiti2}${}^{-}$\cite{Benson}

2. For input data, normalized sample covariance matrices, including amplitude and phase information, are used. Other alternatives include the complex pressure,\cite{Lefort} phase difference,\cite{Steinberg} eigenvalues,\cite{Ozard} amplitude of the pressure field,\cite{Caiti2} transmission loss,\cite{Stephan}${}^{,}$\cite{Benson}  angular dependence of backscatter,\cite{Michalopoulou2} or features extracted from spectrograms.\cite{Thode} This preprocessing procedure is known as feature extraction in machine learning.

3. Under machine learning framework, source localization can be solved as a classification or a regression problem. This work focuses on classification in addition to the regression approach used in previous studies.\cite{Lefort}${}^{,}$\cite{Steinberg}${}^{,}$\cite{Caiti1}${}^{-}$\cite{Benson}

4. Well-developed machine learning libraries are used. Presently, there are numerous efficient open source machine learning libraries available, including TensorFlow,\cite{Tensorflow} Scikit-learn,\cite{Scikit} Theano,\cite{Theano} Caffe,\cite{Caffe} and Torch,\cite{Torch} all of which solve typical machine learning tasks with comparable efficiency. Here, TensorFlow is used to implement FNN because of its simple architecture and wide user base. Scikit-learn is used to implement SVM and RF as they are not included in the current TensorFlow version. Compared to older neural network implementations, Tensorflow includes improved optimization algorithms\cite{Kingma} with better convergence, more robust model with dropout\cite{Srivastava} technique and high computational efficiency. 

The paper is organized as follows. The input data preprocessing and source range mapping are discussed in Secs.~\ref{subsec:2:1} and \ref{subsec:2:2}. The theoretical basis of FNN, SVM and RF is given in Secs.~\ref{subsec:2:3}--\ref{subsec:2:5}. Simulations and experimental results in Secs.~\ref{sec:3} and \ref{sec:4} demonstrate the performance of FNN, SVM and RF. In Sec.~\ref{sec:5}, the effect of varying the model parameters is discussed. The conclusion is given in Sec.~\ref{sec:6}.

\section{\label{sec:2} Localization based on machine learning}
 The dynamics of the ocean and its boundary cause a stochastic relationship between the received pressure phase and amplitude at the array and the source range. After preprocessing we assume a deterministic relationship between ship range and sample covariance matrix. The pressure--range relationship is in general unknown but may be discovered using machine learning methods. The received pressure is preprocessed and used as the input of the machine learning models (Sec.~\ref{subsec:2:1}). The desired output may be either discrete (classification) or continuous (regression) corresponding to the estimated source range (Sec.~\ref{subsec:2:2}). The theory of FNN, SVM, and RF are described in Secs.~\ref{subsec:2:3}--\ref{subsec:2:5}.

\subsection{\label{subsec:2:1} Input data preprocessing}

To make the processing independent of the complex source spectra, the  received array pressure is transformed to a normalized sample covariance matrix. The complex pressure at frequency $f$ obtained by taking the DFT of the input pressure data at $L$ sensors is denoted by $\mathbf{p}(f)=[p_1(f), \cdots, p_L(f)]^T$. The sound pressure is modeled as
\begin{equation} \label{press_model}
\mathbf{p}(f)= S(f)\mathbf{g}(f,{\mathbf{r}})+\mathbf{\epsilon},
\end{equation}
where $\mathbf{\epsilon}$ is the noise, $S(f)$ is the source term, and $\mathbf{g}$ is the Green's function. To reduce the effect of the source amplitude $|S(f)|$, this complex pressure is normalized according to
\begin{equation} \label{press}
\tilde{\mathbf{p}}(f) = \frac{\mathbf{p}(f)}{\sqrt{\sum\limits_{l=1}^{L} \left |{p_{l}(f)}\right |^2 }} = \frac{\mathbf{p}(f)}{\|\mathbf{p}(f)\|_2}.
\end{equation}

The normalized sample covariance matrices (SCMs) are averaged over $N_s$ snapshots to form the conjugate symmetric matrix
\begin{equation} \label{SCM}
\mathbf{C}(f) = \frac{1}{N_s} \sum_{s=1}^{N_s} \tilde{\mathbf{p}}_s(f)  \tilde{\mathbf{p}}_s^{H}(f),
\end{equation}
where $H$ denotes conjugate transpose operator and $\tilde{\mathbf{p}}_s$ represents the sound pressure over the {\it s}th snapshot. The product $\tilde{\mathbf{p}}_s(f)  \tilde{\mathbf{p}}_s^{H}(f)$ contains an $S(f) S(f)^H$ term, which for large SNR is dominant and thus reduces the effect of the source phase. Preprocessing the data according to Eqs.~(\ref{press}) and (\ref{SCM}) ensures that the Green's function is used for localization. Only the real and imaginary parts of the complex valued entries of diagonal and upper triangular matrix in $\mathbf{C}(f)$ are used as input to save memory and improve calculation speed. These entries are vectorized to form the real-valued input $\mathbf{x}$ of size $L \times (L+1)$ to the FNN, SVM and RF. 

\subsection{\label{subsec:2:2} Source range mapping}

In the classification problem, a set of source ranges is discretized into $K$ bins, $r_1,...,r_K$, of equal width $\Delta r$. Each input vector, $\mathbf{x}_n, n=1,..,N$, is labeled by $t_n$, where $t_n \in r_k, k = 1,...,K$; this label represents the true source range class and is the target output for the model. SVM and RF use this classification scheme to train and predict the source range for each sample.

For the FNN, the range class $t_n$ is mapped to a $1\times K$ binary vector, $\mathbf{t}_n$, such that:
\begin{equation}
t_{nk} = \left\{ \begin{array} {ll} 1 & \quad \mbox{if } |t_n-r_k| \leq \frac{\Delta r}{2},\\
0 & \quad \mbox{otherwise, }\end{array}
\right.
\end{equation}
$\mathbf{t}_n=t_{n,1},...,t_{n,K}$ therefore represents the expected output probability of the neural network, i.e. the probability that the source is at range $r_k$ for input $\mathbf{x}_n$. These target vectors are used to train the FNN. The FNN output predictions are given as a softmax distribution with maximum at the predicted range (see Sec. \ref{subsec:2:3}).

In the regression problem, the target output $r_n \in [0, \infty)$ is a continuous range variable for all three models. 

\subsection{\label{subsec:2:3} Feed-forward neural networks}

The feed-forward neural network (FNN), also known as multi-layer perceptron, is constructed using a feed-forward directed acyclic architecture, see Fig.~\ref{fig:FIG1}(a). The outputs are formed through a series of functional transformations of the weighted inputs. In the FNN, the outputs are deterministic functions of the inputs.\cite{Bishop} 

\begin{figure}
\centering
\includegraphics{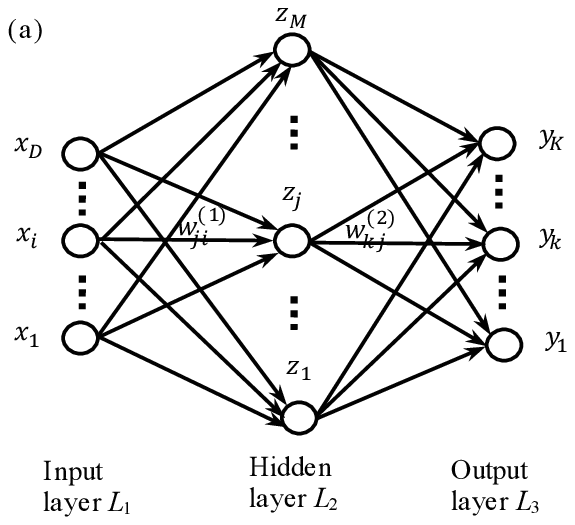}
\includegraphics{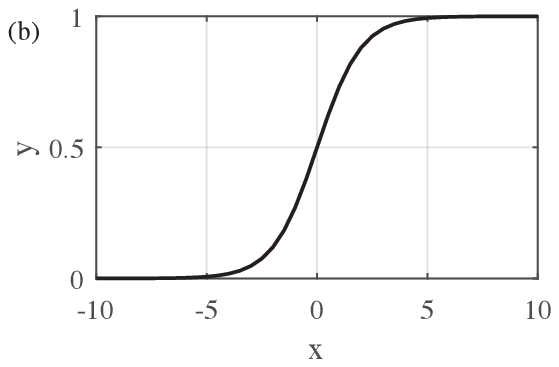}
\caption{ \label{fig:FIG1} (a) Diagram of a feed-forward neural network and (b) Sigmoid function.}
\end{figure}

Here, a three layer model (input layer $L_1$, hidden layer $L_2$ and output layer $L_3$) is used to construct the FNN. The input layer $L_1$ is comprised of $D$ input variables $\mathbf{x}=[x_1,\cdots,x_D]^T$. The $j$th linear combination of the input variables is given by
\begin{equation} \label{weighted sum1}
a_j= \sum_{i=1}^{D} w_{ji}^{(1)}x_{i} + w_{j0}^{(1)},\qquad j=1,\cdots,M,
\end{equation}
where $M$ is the number of neurons in $L_2$ and the superscript indicates that the corresponding parameters are in the first layer of the network. The parameters $w_{ji}^{(1)}$ and $w_{j0}^{(1)}$ are called the weights and biases and their linear combinations $a_j$ are called activations. In $L_2$, the activations are transformed using an activation function $f(\cdot )$,
\begin{equation} \label{output}
z_j = f(a_j).
\end{equation}
The logistic sigmoid was chosen as the intermediate activation function for this study, see Fig.~\ref{fig:FIG1}(b):
\begin{equation} \label{sigmoid}
f(a) = \sigma(a)=\frac{1}{1+e^{-a}}.
\end{equation}

Similarly, for output layer $L_3$, the $K$ output unit activations are expressed as linear combinations of $z_j$
\begin{equation} \label{weighted sum2}
a_k = \sum_{j=1}^{M} w_{kj}^{(2)} z_j + w_{k0}^{(2)},  \qquad k=1,\cdots,K
\end{equation}
where $w_{kj}^{(2)}$ and $w_{k0}^{(2)}$ represent weights and biases for the second layer. 

In the output layer, the softmax function is used as the activation function. The softmax is a common choice for multi-class classification problems.\cite{Bishop} Here, it constrains the output class, $y_k(\mathbf{x},\mathbf{w})$, to be the probability that the source is at range $r_k$:\cite{Bishop}
\begin{equation} \label{softmax}
y_k(\mathbf{x},\mathbf{w}) = \frac{\exp(a_k(\mathbf{x},\mathbf{w}))}{\sum_{j=1}^{K}{\exp({a_j(\mathbf{x},\mathbf{w}))}}}, \quad k=1,\cdots,K
\end{equation}
where $\mathbf{w}$ is the set of all weight and bias parameters and $y_k$ satisfies $0\le y_k\le 1$ and $\sum_{k}{y_k=1}$.

Before applying the FNN to unlabeled data, the weights and biases $\mathbf{w}$ are determined by training the model on labeled data. Recall that in the FNN case, $\mathbf{t}_n$ is the binary target vector, or true probability distribution (see Sec.~\ref{subsec:2:2}), and $y_k(\mathbf{x}_n,\mathbf{w})$ is the estimated probability distribution, for the input $\mathbf{x}_n$ (see Sec.~\ref{subsec:2:1}).

During training, the Kullback--Leibler (KL) divergence
\begin{equation} \label{KL}
D_\text{KL}(\mathbf{t}_n||\mathbf{y}(\mathbf{x}_n,\mathbf{w}))=\sum_{k} t_{nk} \left[ \ln t_{nk} -\ln y_{nk} \right],  
\end{equation}
represents the dissimilarity between $y_{nk} = y_k(\mathbf{x}_n,\mathbf{w})$ and $t_{nk}$, where $\mathbf{t}_n=\left[ t_{n1}, ... , t_{nk}\right]$, $k = 1, ... , K$. Minimizing the KL divergence $D_\text{KL}$ is equivalent to minimizing the cross entropy function $E_n$ 
\begin{equation} \label{cross_entropy}
E_{n}(\mathbf{t}_n,\mathbf{y}(\mathbf{x}_n,\mathbf{w}))=-\sum_{k} t_{nk}\ln y_{nk},
\end{equation}
since the desired output $\mathbf{t}_n$ is constant (independent of $\mathbf{w}$).

For $N$ observation vectors, the averaged cross entropy and resulting weights and biases are
\begin{equation}  \label{cross_entropy2} 
E(\mathbf{w})=-\frac{1}{N}\sum_{n=1}^{N} \sum_{k=1}^{K} t_{nk} \ln y_{nk}, 
\end{equation}
\begin{equation}  \label{weights} 
\mathbf{\hat{w}}= \underset{\mathbf{w}}{\operatorname{argmin}} \left[ - \frac{1}{N}\sum_{n=1}^{N} \sum_{k=1}^{K} t_{nk} \ln y_{nk}\right].
\end{equation}

For the regression problem, there is only one neuron in the output layer representing the continuous range variable. Instead of using Eq. (\ref{cross_entropy2}), a sum-of-squares error function\cite{Bishop} is minimized
\begin{equation}  \label{FNN_regression} 
E(\mathbf{w})=\frac{1}{2}\sum_{n=1}^{N} \left | y(\mathbf{x}_n,\mathbf{w})-r_n \right | ^2,
\end{equation}
where $r_n$ is the true source range at sample $n$.

Several optimization methods are provided in the TensorFlow software. In this paper, Adam\cite{Kingma}(Adaptive Moment estimation) is used.

\subsection{\label{subsec:2:4} Support Vector Machine}

Unlike neural networks, support vector machines (SVM) are decision machines that do not provide a posterior probability.\cite{Bishop} Instead, the data is divided into two (or more) classes by defining a hyperplane that maximally separates the classes.

\begin{figure}
\includegraphics{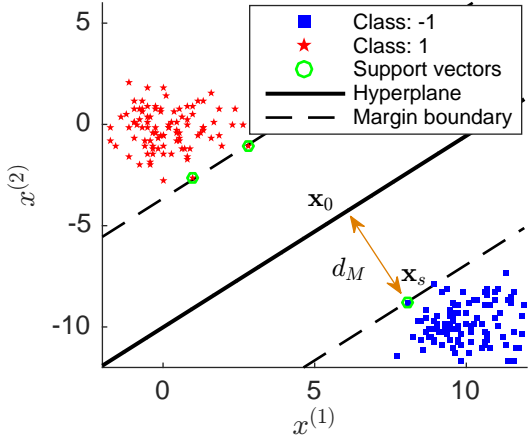}
\caption{\label{fig:FIG2}{(Color online) A linear hyperplane learned by training an SVM in two dimensions ($D=2$).}}
\end{figure}

First, for simplicity, assume the input $\mathbf{x}_n, n = 1,\cdots, N$ are linearly separable (see Fig.~\ref{fig:FIG2}) and can be divided into two classes, $s_n \in \{1, -1\}$. The class of each input point $\mathbf{x}_n$ is determined by the form\cite{Bishop}
\begin{equation}\label{relation}
\text{y}_n = \mathbf{w}^T \mathbf{x}_n + b,
\end{equation}
where $\mathbf{w}$ and $b$ are the unknown weights and bias. A hyperplane satisfying $\mathbf{w}^T \mathbf{x} + b=0$ is used to separate the classes. If $\text{y}_n$ is above the hyperplane ($\text{y}_n>0$), estimated class label $\hat{s}_{n}=1$ , whereas if $\text{y}_n $ is below ($\text{y}_n<0$), $\hat{s}_{n}=-$1. The perpendicular distance $d$ of a point ${\mathbf{x}}_n$ to the hyperplane is the distance between the point $\mathbf{x}_n$ and its projection $\mathbf{x}_0$ on the hyperplane, satisfying
\begin{equation}\label{d1}
\begin{aligned}
&\mathbf{x}_n=\mathbf{x}_0+d \frac{\mathbf{w}}{||\mathbf{w}||}, \\
&\mathbf{w}^T \mathbf{x}_0 + b=0, 
\end{aligned}
\end{equation}
where $||\cdot||$ is the $l_2$ norm. From Eq.~(\ref{d1}), the distance $d$ is obtained:
\begin{align}
d(\mathbf{x}_n)=s_n\frac{\mathbf{w}^T \mathbf{x}_n + b}{||\mathbf{w}||}, \label{d}
\end{align}
where $s_n$ is added in Eq.~(\ref{d}) to guarantee $d > 0$. The margin $d_M$ is defined as the distance from the hyperplane to the closest points $\mathbf{x}_s$ on the margin boundary (support vectors, see Fig.~\ref{fig:FIG2}). The optimal $\mathbf{w}$ and $b$ are solved by maximizing the margin $d_M$:
\begin{equation} \label{svm_max1}
\begin{aligned}
 &\underset{\mathbf{w}, b}{\operatorname{argmax}} \quad d_M,  \\
\text{subject to  }  & \frac{s_n (\mathbf{w}^T \mathbf{x}_n + b)}{||\mathbf{w}||} \geq d_M, \quad n=1,\cdots, N.
\end{aligned}
\end{equation}
The Eq.~(\ref{svm_max1}) is equivalent to this optimization problem:\cite{Bishop}
\begin{equation} \label{svm_min}
\begin{aligned}
&\underset{\mathbf{w}, b}{\operatorname{argmin}} \quad \frac{1}{2} ||\mathbf{w}||^2,\\
\text{subject to  } & s_n (\mathbf{w}^T \mathbf{x}_n + b) \geq 1,  \quad n=1,\cdots, N. 
\end{aligned}
\end{equation}

If the training set is linearly non-separable (class overlapping), slack variables\cite{Bishop} $\xi_n \geq 0$ are introduced to allow some of the training points to be miclassified, corresponding the optimization problem:
\begin{equation} \label{svm_min2}
\begin{aligned}
&\underset{\mathbf{w}, b}{\operatorname{argmin}} \quad \frac{1}{2} ||\mathbf{w}||^2 + C \sum_{n=1}^{N}{\xi_n},\\
\text{subject to  } & s_n \text{y}_n  \geq 1-\xi_n,  \quad n=1,\cdots, N. 
\end{aligned}
\end{equation}
The parameter $C>0$ controls the trade-off between the slack variable penalty and the margin. 

Often the relation between $\text{y}_n$ and $\mathbf{x}_n$ is nonlinear. Thus Eq.~(\ref{relation}) becomes
\begin{equation} \label{non_model}
\text{y}_n = \mathbf{w}^T \phi(\mathbf{x}_n) + b,
\end{equation}
where $\phi(\mathbf{x}_n)$ denotes the feature-space transformation. By substituting $\phi(\mathbf{x}_n)$ for $\mathbf{x}_n$, Eqs.~(\ref{d1}--\ref{svm_min2}) are unchanged.

The constrained optimization problem can be rewritten in terms of the dual Lagrangian form:\cite{Bishop}
\begin{equation} \label{dual}
\begin{aligned}
\tilde{L}(\mathbf{a})  =  \sum_{n=1}^{N}{a_n}&-\frac{1}{2}\sum_{n=1}^{N}\sum_{m=1}^{N} a_n a_m s_n s_m k_{\phi}(\mathbf{x}_n,\mathbf{x}_m), \\
 \text{subject to  }  0 & \leq a_n \leq C, \\
  &\sum_{n=1}^{N}{a_n s_n} = 0, \quad n=1,\cdots, N, 
\end{aligned}
\end{equation}
where $a_n \geq 0$ are Lagrange multipliers and dual variables, and $k_{\phi}(\mathbf{x}_n,\mathbf{x}_m) =  \phi(\mathbf{x}_n)^T \phi(\mathbf{x}_m)$ is the kernel function. In this study, we use the Gaussian radial basis function (RBF) kernel\cite{Scikit}
\begin{equation} \label{RBF}
k_{\phi}(\mathbf{x}, \mathbf{x'}) = \exp(-\gamma ||\mathbf{x} - \mathbf{x'}||^2),
\end{equation}
where $\gamma$ is a parameter that controls the kernel shape.

Support vector regression (SVR) is similar to SVM, but it minimizes the $\epsilon$--sensitive error function
\begin{equation}
\mathcal{E}_\epsilon(\text{y}_n - r_n) = 
\begin{cases}
0, & \text{ if  } |\text{y}_n - r_n| < \epsilon, \\
|\text{y}_n-r_n| - \epsilon, & \text{ otherwise},
\end{cases}
\end{equation}
where $r_n$ is the true source range at sample $n$ and $\epsilon$ defines a region on either side of the hyperplane. In SVR, the support vectors are points outside the $\epsilon$ region. 

Because the SVM and SVR models are a two-class models, multi-class SVM with $K$ classes is created by training $K(K-1)/2$ models on all possible pairs of classes. The points that are assigned to the same class most frequently are considered to comprise a single class, and so on until all points are assigned a class from $1$ to $K$. This approach is known at the $^{``}$one-versus-one" scheme,\cite{Bishop} although slight modifications have been introduced to reduced computational complexity.\cite{Fan}

\subsection{\label{subsec:2:5} Random forests}

The random forest (RF)\cite{Breiman1} classifier is a generalization of the decision tree model, which greedily segments the input data into a predefined number of regions. The simple decision tree model is made robust by randomly training subsets of the input data and averaging over multiple models in RF.

\begin{figure}
\includegraphics{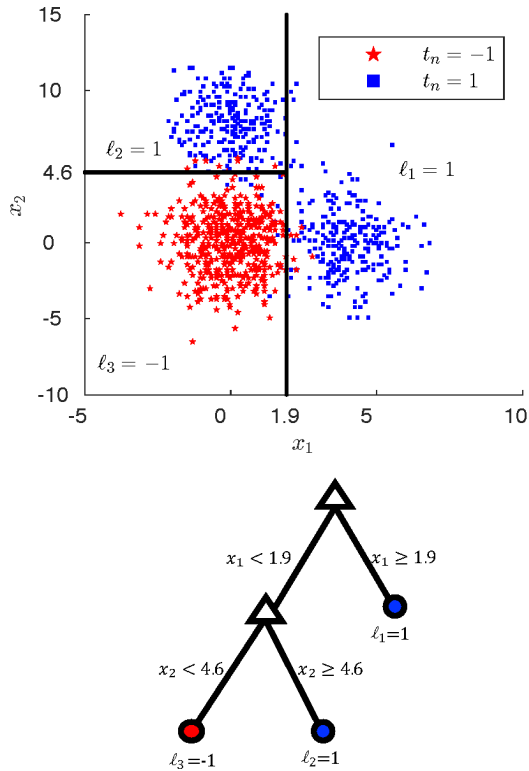}
\caption{ \label{fig:FIG3}{(Color online) Decision tree classifier and corresponding rectangular regions shown for two--dimensional data with $K=2$ classes ($D=2$, $M=3$) and 1000 training points.}}
\end{figure}

Consider a decision tree (see Fig.~\ref{fig:FIG3}) trained on all the input data. Each input sample, $\mathbf{x}_n, n = 1,...,N$, represents a point in $D$ dimensions. The input data can be partitioned into two regions by defining a cutoff along the $i$th dimension, where $i$ is the same for all input samples $\mathbf{x}_n, n = 1,...,N$:
\begin{equation}
\begin{aligned}
\mathbf{x}_n &\in \bm{x}_\text{left}   \quad & \text{ if }  x_{ni} > c,  \\
\mathbf{x}_n &\in \bm{x}_\text{right}  \quad &\text{ if } x_{ni} \leq c.
\end{aligned}
\end{equation}
$c$ is the cutoff value, and $\bm{x}_\text{left}$ and $\bm{x}_\text{right}$ are the left and right regions, respectively. The cost function, $G$, that is minimized in the decision tree at each branch is\cite{Scikit}
\begin{equation}
\begin{aligned}
c^* &= \underset{c}{\operatorname{argmin }}  \text{ } G(c), \\
G(c) &= \frac{n_\text{left}}{N} H(\bm{x}_\text{left}) + \frac{n_\text{right}}{N}H(\bm{x}_\text{right}),
\end{aligned}
\end{equation}
where $n_\text{left}$ and $n_\text{right}$ are the numbers of points in the regions $\bm{x}_\text{left}$ and $\bm{x}_\text{right}$. $H(\cdot)$ is an impurity function chosen based on the problem.

For the classification problem, the Gini index\cite{Scikit} is chosen as the impurity function
\begin{equation}
H(\bm{x}_m) = \frac{1}{n_m}\sum_{\mathbf{x}_n \in \bm{x}_m} I(t_n, \ell_m) \left[1 - \frac{1}{n_m}I(t_n, \ell_m) \right],
\end{equation}
where $n_m$ is the number of points in region $\bm{x}_m$ and $\ell_m$ represents the assigned label for each region, corresponding to the most common class in the region:\cite{Scikit}
\begin{equation} \label{region_label}
\ell_m= \underset{r_k}{\operatorname{argmax}} \text{ }\sum_{\mathbf{x}_n \in \bm{x}_m} I(t_n, r_k).
\end{equation}
In Eq.~(\ref{region_label}), $r_k, k = 1,...,K$ are the source range classes and $t_n$ is the label of point $\mathbf{x}_n$ in region $m$, and 
\begin{equation}
I(t_n, r_k) = 
\begin{cases}
1 & \text{ if } t_n = r_k, \\
0 & \text{ otherwise.}
\end{cases}
\end{equation}

The remaining regions are partitioned iteratively until regions $\bm{x}_1,...,\bm{x}_M$ are defined. In this paper, the number of regions, $M$, is determined by the minimum number of points allowed in a region.  A diagram of the decision tree classifier is shown in Fig.~\ref{fig:FIG3}. The samples are partitioned into $M=3$ regions with the cutoff values 1.9 and 4.6.

For RF regression, there are two differences from classification: the estimated class for each region is defined as the mean of the true class for all points in the region, and the mean squared error is used as the impurity function
\begin{equation}
\begin{aligned}
\ell_m &= \frac{1}{n_m}\sum_{\mathbf{x}_n \in \bm{x}_m} r_n, \\
H(\bm{x}_m) &= \sum_{\mathbf{x}_n \in \bm{x}_m} (\ell_m - r_n)^2,
\end{aligned}
\end{equation}
where $r_n$ is source range at sample $n$.

As the decision tree model may overfit the data, statistical bootstrap and bagging are used to create a more robust model, a random forest.\cite{Breiman} In a given draw, the input data,  $\mathbf{x}_i, i = 1,\cdots,Q$, is selected uniformly at random from the full training set, where $Q \leq N$. $B$ such draws are conducted with replacement and a new decision tree is fitted to each subset of data. Each point, $\mathbf{x}_n$, is assigned to its most frequent class among all draws:
\begin{equation}
\hat{f}^\text{bag}(\mathbf{x}_n) = \underset{t_n}{\operatorname{argmax}}\sum_{b=1}^B I(\hat{f}^{\text{tree},b}(\mathbf{x}_n), t_n),
\end{equation}
where $\hat{f}^{\text{tree},b}(\mathbf{x}_i)$ is the class of $\mathbf{x}_i$ for the $b$th tree.

\subsection{\label{subsec:2:6} Performance metric}

To quantify the prediction performance of the range estimation methods, the mean absolute percentage error (MAPE) over $N$ samples is defined as
\begin{equation} \label{MAPE}
 E_\text{MAPE} = \frac{100}{N}\sum_{i=1}^{N} \left | \frac{Rp_i-Rg_i}{Rg_i} \right |,
\end{equation}
where $Rp_i$ and $Rg_i$ are the predicted range and the ground truth range, respectively. MAPE is preferred as an error measure because it accounts for the magnitude of error in faulty range estimates as well as the frequency of correct estimates. MAPE is known to be an asymmetric error measure\cite{Goodwin} but is adequate for the small range of outputs considered. 

\subsection{\label{subsec:2:7} Source localization algorithm}

The localization problem solved by machine learning is implemented as follows:

1. Data preprocessing. The recorded pressure signals are Fourier transformed and $N_s$ snapshots form the SCM from which the input ${\mathbf x}$ is formed. 

2. Division of preprocessed data into training and test data sets. For the training data, the labels are prepared based on different machine learning algorithms.

3. Training the machine learning models. $\mathbf{X}=[\mathbf{x}_1 \cdots \mathbf{x}_N]$ are used as the training input and the corresponding labels as the desired output. 

4. Prediction on unlabeled data. The model parameters trained in step 3 are used to predict the source range for test data. The resulting output is mapped back to range, and the prediction error is reported by the mean absolute percentage error.

\section{\label{sec:3}  Simulations}

In this section, the performance of machine learning on simulated data is discussed. For brevity, only the FNN classifier is examined here, although the conclusions apply to SVM and RF. Further discussion of SVM and RF performance is included in Secs.~\ref{sec:4} and \ref{sec:5}.

\subsection{\label{subsec:3:1} Environmental model and source-receiver configuration}

Acoustic data is simulated using KRAKEN\cite{Porter} with environmental parameters simulating the Noise09 experiment\cite{Byun}, see Fig.~\ref{fig:FIG4}(a). The source frequency is 300 Hz. The source depth is 5 m in a 152 m waveguide, with a 24 m sediment layer (sound speed 1572--1593 m/s, density 1.76 $\mathrm{g/cm^3}$, attenuation coefficient 2.0 $\mathrm{dB/\lambda}$) and a fluid halfspace bottom (sound speed 5200 m/s, density 1.8 $\mathrm{g/cm^3}$, and attenuation coefficient 2.0 $\mathrm{dB/\lambda}$). The sound speed profile of water column is shown in Fig.~\ref{fig:FIG4}(b). The vertical array consists of 16 receivers spanning 128--143 m depth with inter-sensor spacing 1 m.

A source traveling away from the receiver at 2 m/s is simulated by varying the range from 0.1 to 2.86 km at 2 m intervals. Realizations with different SNRs are generated by adding appropriate complex Gaussian noise to the simulated received complex pressure signals.

Since the source moves in range and the source level is assumed constant, SNR is defined at the most distant range bin
\begin{equation} \label{SNR}
\mathrm{SNR} = 10\log_{10} \frac{\sum_{l=1}^{L} |\hat{p}_l|^2/L}{\sigma ^2} (\mathrm{dB}),
\end{equation}
where $\hat{p}_l$ is sound pressure signal received by the $l$th sensor at the longest source-receiver distance and $\sigma ^2$ represents the noise variance.

\begin{figure}
\begin{center}
\includegraphics{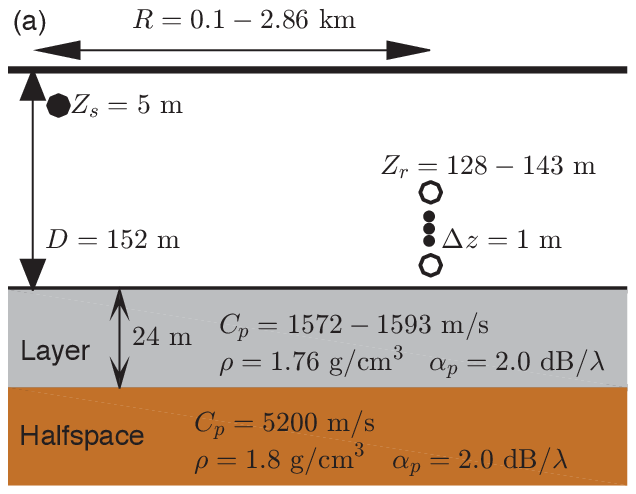}
\includegraphics{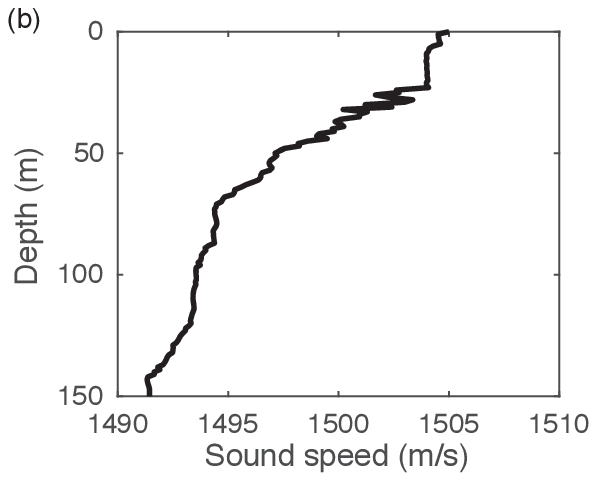}
\caption{ \label{fig:FIG4}{(Color online) (a) Waveguide parameters and source-receiver configuration. (b) Sound speed profile of water column.}}
\end{center}
\end{figure}

\subsection{\label{subsec:3:2} Input preprocessing and learning parameters}

The SCM for a 16--element vertical array is formed at each range point by averaging over $N_s=10$ successive snapshots (9 snapshots overlapped) according to Eq.~(\ref{SCM}). The number of neurons in the input layer is therefore $D=16 \times (16 + 1)=272$.  The range sample interval is 2 m, with 1380 total range samples (1 s duration per snapshot). Thus, a total of $N=1380$ input matrices constitute the sample set spanning the whole range 0.1--2.86 km.

For each SNR, two realizations of noisy measurements are generated. One realization of size 1380 $\times$ 272 is used for the training set. For the test set, the range sample interval is changed to 20 m, and a realization of size 138 $\times$ 272 is used as input.

In the test set, $K=138$ output neurons represent ranges from 0.1--2.86 km incremented by 20 m. The number of neurons in the hidden layer is $M=128$. To prevent overfitting, the "keep dropout" technique,\citep{Srivastava} with probability 0.5, is used. The initial learning rate for the Adam optimizer\citep{Kingma} is 0.01 and the maximum number of iterations is 1000.

\subsection{\label{subsec:3:3} Results}

The prediction performance is examined for four SNRs ($-$10, $-$5, 0, 5 dB). Figure \ref{fig:FIG5} compares range predictions by FNN and the true ranges on test data. For the four SNRs tested, the MAPE for the FNN predictions is 20.6, 6.5, 0.2 and 0.0\%, respectively. 
\begin{figure}
\begin{center}
\includegraphics{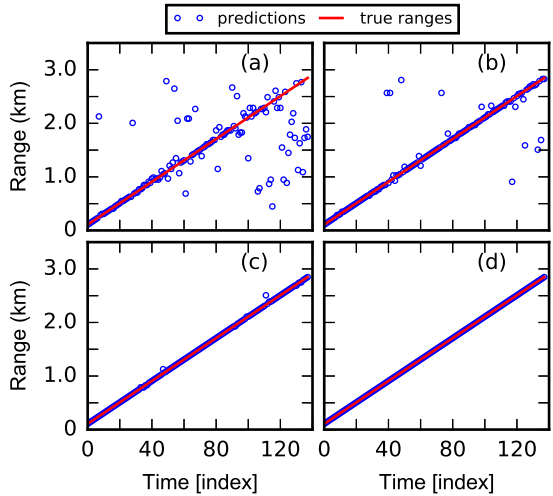}
\caption{ \label{fig:FIG5}{(Color online) Range predictions by FNN on test data set with SNR of (a) $-$10, (b) $-$5, (c) 0, and (d) 5 dB. The time index increment  is 10 s.}}
\end{center}
\end{figure}

As described in Sec.~\ref{sec:2}, the output $y_{nk}$ of FNN represents the probability distribution over a discrete set of possible ranges. To demonstrate the evolution of the probability distribution as the FNN is trained, $y_{nk}$ versus training steps is plotted in Fig.~\ref{fig:FIG6} for the signal with SNR 5 dB at range 1.5 km. After 300 training steps, the FNN output probability distribution resembles the target output.

\begin{figure}
\begin{center}
\includegraphics{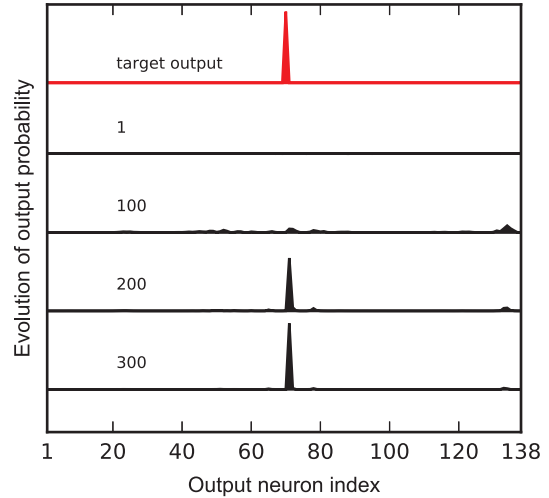}
\caption{ \label{fig:FIG6} (Color online) Output probability for range 0.1--2.86 km (the true range is 1.5 km) after training steps (1, 100, 200, 300). The top line represents the target output.}
\end{center}
\end{figure}

In Fig. \ref{fig:FIG7}, the convergence of the FNN algorithm is investigated by plotting the cross entropy Eq.~(\ref{cross_entropy2}) versus the optimization step on training and test data. It shows that the FNN converges after about 300 steps at all SNRs. For low SNRs ($<0$ dB), the FNN classifier generates poor predictions on test data while performing well on training data, which indicates overfitting.

Increasing the training set size can reduce overfitting but additional data may be not available due to experimental or computational constraints. For higher SNRs (e.g., 0 and 5 dB), both test and training errors converge to low cross entropy, indicating good performance. Therefore, best performance of machine learning methods is expected for high SNR.

\begin{figure}
\begin{center}
\includegraphics{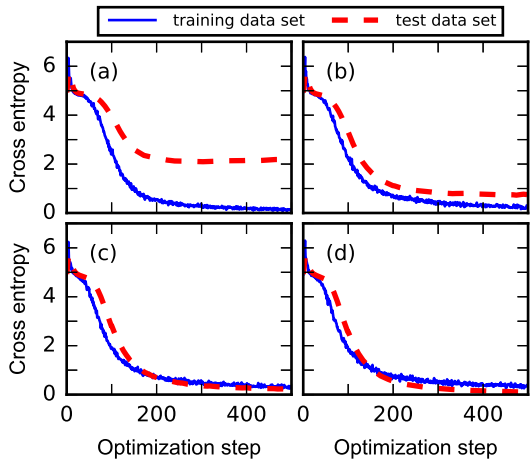}
\caption{\label{fig:FIG7}{(Color online) Cross entropy Eq.~(\ref{cross_entropy}) versus optimization steps on training (solid) and test (dashed) data with SNR of (a) $-$10, (b) $-$5, (c) 0, and (d) 5 dB.}}
\end{center}
\end{figure}

\begin{figure}
\begin{center}
\includegraphics{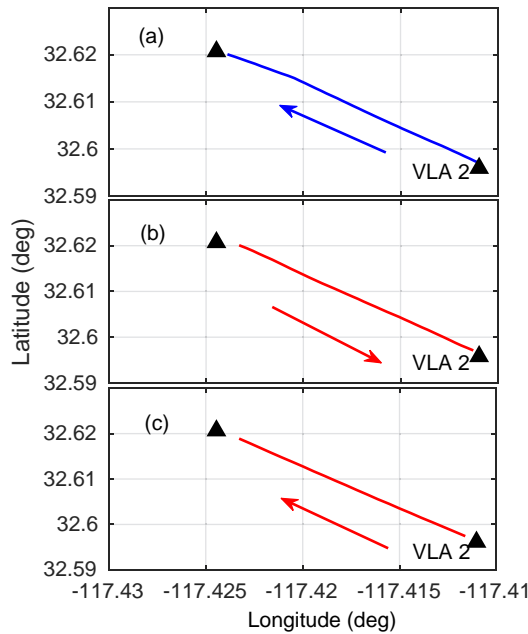}
\caption{ \label{fig:FIG8}{(Color online) Ship tracks for Noise09 experiment during the periods (a) 01/31/2009, 01:43--02:05 (training data, ship speed 2 m/s), (b) 01/31/2009, 01:05--01:24 (Test-Data-1, ship speed $-$2 m/s), and (c) 02/04/2009, 13:41--13:51 (Test-Data-2, ship speed 4 m/s).}}
\end{center}
\end{figure}

\section{\label{sec:4} Experimental results}

Shipping noise data radiated by R/V New Horizon during the Noise09 experiment are used to demonstrate the performance of the FNN, SVM and RF localization. The experiment geometry is shown in Fig.~\ref{fig:FIG8}, with bottom-moored vertical linear arrays (VLAs) indicated by triangles and the three ship tracks used for range estimation. The hydrophone sampling rate was 25 kHz.

The data from VLA2, consisting of 16 hydrophones at 1 m spacing, are used for range estimation. The frequency spectra of shipping noise recorded on the top hydrophone during the three periods are shown in Fig.~\ref{fig:FIG9}. The striations  indicate that the source was moving. The SNR decreases with increasing source-receiver distance.

Data from period 01:43--02:05 on January 31, 2009 are used as the training set and 01:05--01:24 on January 31 and 13:41--13:51 on February 4 are used as the test sets (Test-Data-1 and Test-Data-2). 

The GPS antenna on the New Horizon is separated from the noise--generating propeller by a distance $L_d$. To account for this difference we use the range between the propeller and VLA2 as the ground truth range $R_g$:\begin{equation} \label{correction}
 R_g =\left \{
 \begin{aligned}
 &R_\text{GPS} - L_d  \quad  \rm{for~training~data~and~Test\text{-}Data\text{-}2}, \\
 &R_\text{GPS} + L_d  \quad  \rm{for~Test\text{-}Data\text{-}1},
  \end{aligned}
  \right.
 \end{equation}
where $R_{GPS}$ represents the range between the GPS antenna and VLA2. According to the  R/V New Horizon handbook, $L_d=24.5$ m. In the following, the ranges have been corrected by Eq.~(\ref{correction}).

\begin{figure}
\begin{center}
\includegraphics{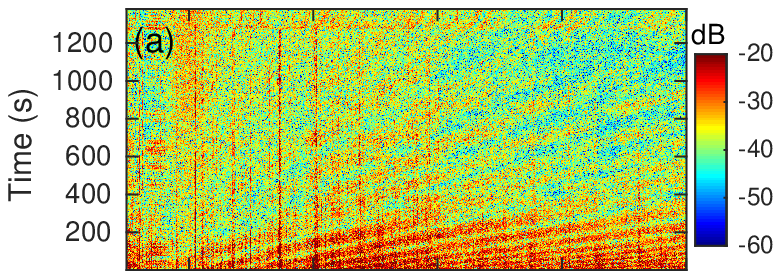}\\
\includegraphics{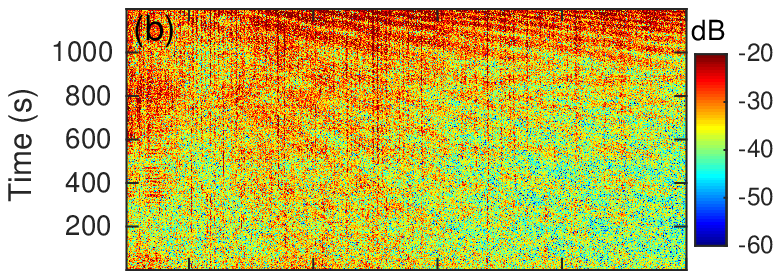}\\
\includegraphics{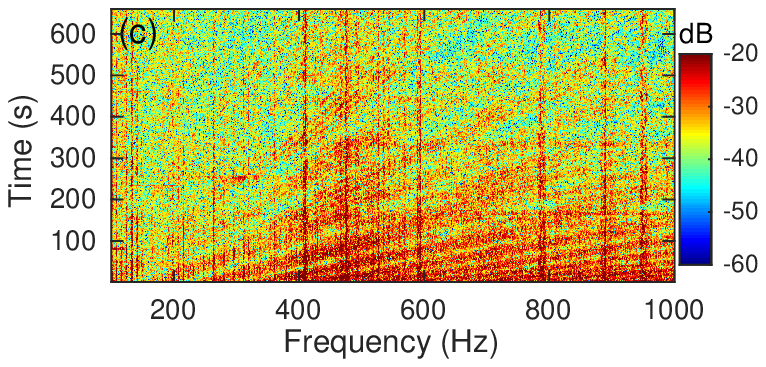}
\caption{ \label{fig:FIG9} {(Color online) Spectra of shipping noise during periods (a) 01/31/2009, 01:43--02:05, (b) 01/31/2009, 01:05--01:24, and (c) 02/04/2009, 13:41--13:51. }}
\end{center}
\end{figure}

\subsection{\label{subsec:4:1} Input preprocessing and learning parameters}

For the training set and both test sets, the $16 \times 16$ SCM at each range and frequency, averaged over 10 successive 1-s snapshots, is used as input. There are 1380 samples in the training data set and 120 samples in each of the test data sets (samples are drawn every 10 s for Test-Data-1 and 5 s for Test-Data-2). The source-receiver range 0.1--3 km is divided into $K=138$ discrete range points.

As in the simulations in Sec.~\ref{subsec:3:2}, the keep probability for training dropout of the FNN is 0.5, the initial learning rate is 0.01 and the maximum iteration step is 1000. The number of neurons in the hidden layer is chosen as $M=128$ for 1 frequency and $M=1024$ for 66 frequencies ( see Sec.~\ref{subsec:4:2}). 

For the SVM classifier, Gaussian radial basis function kernel is used. The parameters $\gamma$ (Eq.~(\ref{RBF})) and $C$ (Eq.~(\ref{svm_min2})) were tested over $[10^{-3} \quad 10^{-1}]$ and $[10 \quad 10^3]$, respectively. Values of $\gamma=10^{-2}$ and $C=10$ are found to be optimal.

For the RF method, the number of trees bagged is 500, with a minimum of 50 samples required for each leaf.

The performance of all test cases for the FNN, SVM, RF and conventional MFP is summarized in Tables~\ref{table:1} and \ref{table:2}.

\subsection{\label{subsec:4:2} SCM inputs}

Because the shipping noise has a wide frequency band as seen from Fig.~\ref{fig:FIG9}, the performance of the machine learning with the single  and multi-frequency inputs is investigated. The FNN classifier is again used an example to illustrate the benefit of using multiple frequencies.

Input SCMs are formed at 550, 950, and 300--950 Hz with 10 Hz increments (66 frequencies). For the multi-frequency input, the SCMs are formed by concatenating multiple single-frequency SCM input vectors. For example, the dimension of a single frequency input sample is 272, whereas the multi-frequency input has a dimension $272 \times N_f$ for $N_f$ frequencies. The FNN is trained separately for each case and the source-receiver range is then predicted at the selected frequencies.

The prediction results on the two test data sets are shown in Figs.~\ref{fig:FIG10}(a--f) along with $R_g$. For single frequency inputs, the minimum error is 12\% (Fig.~\ref{fig:FIG10}(d)) at 550 Hz and the largest error is 18\% at 950 Hz (Fig.~\ref{fig:FIG10}(e)), both on Test-Data-2. For multi-frequency inputs, the prediction error is 8\% on Test-Data-1 and 6\% on Test-Data-2, indicating a performance improvement by using multiple frequencies. In general, the FNN predictions are better at close ranges due to higher SNR, as expected from the simulation results. However, the FNN with multi-frequency inputs performs well regardless of source range.

\begin{figure}
\begin{center}
\includegraphics{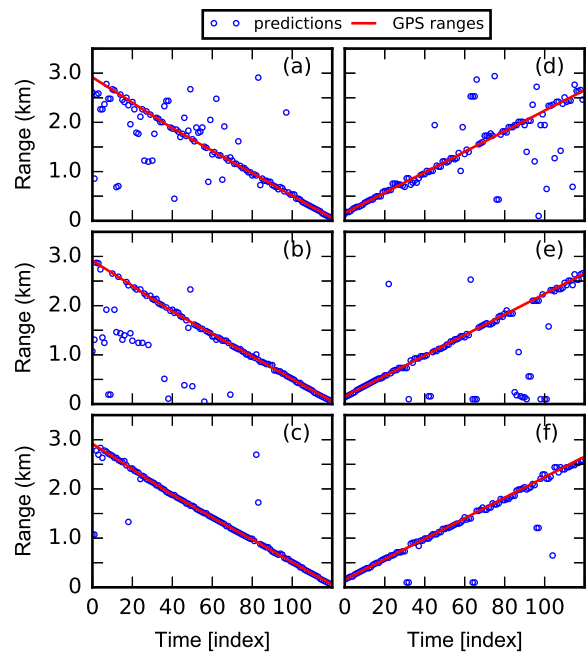}
\caption{ \label{fig:FIG10} (Color online) Range predictions on Test-Data-1 (a, b, c) and Test-Data-2 (d, e, f) by FNN. (a)(d) 550 Hz,  (b)(e) 950 Hz,  (c)(f) 300--950 Hz with 10 Hz increment, i.e. 66 frequencies. The time index increment  is 10 s for Test-Data-1, and 5 s for Test-Data-2.}
\end{center}
\end{figure}

\subsection{\label{subsec:4:3} Source localization as a classification problem}

Source localization is first solved as a classification problem. Only the best MAPE obtained by FNN (Sec.~\ref{subsec:2:3}), SVM (Sec.~\ref{subsec:2:4}) and RF (Sec.~\ref{subsec:2:5}) is shown here (Fig.~\ref{fig:FIG11}). These results are summarized in Table~\ref{table:1}. 

The lowest MAPE is achieved by the SVM, with 2\% on both data sets. RF also reaches 2\% MAPE for Test-Data-2 and 3\% for Test-Data-1. FNN has 3\% MAPE for both test sets. The performance of these three machine learning algorithms is comparable when solving range estimation as a classification problem.

The performance of these machine learning algorithms with various parameters (e.g. number of classes, number of snapshots and model hyper-parameters) is examined in Sec.~\ref{sec:5}.

\begin{figure}
\begin{center}
\includegraphics{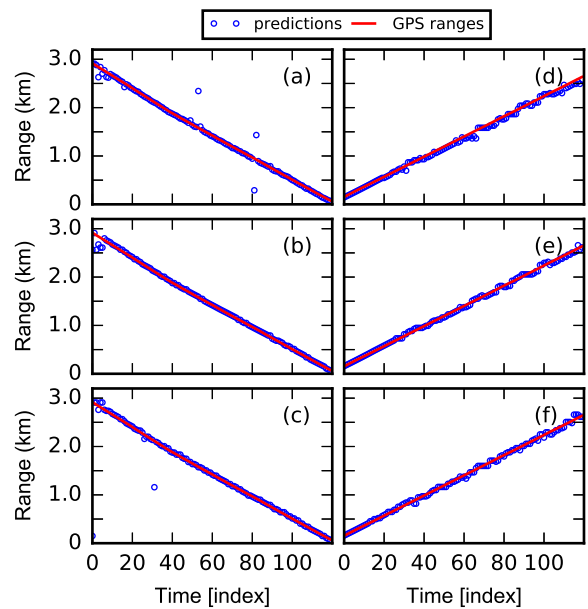}
\caption{ \label{fig:FIG11} {(Color online) Source localization as a classification problem. Range predictions on Test-Data-1 (a, b, c) and Test-Data-2 (d, e, f) by FNN, SVM and RF for 300--950 Hz with 10 Hz increment, i.e. 66 frequencies. (a)(d) FNN classifier, (b)(e) SVM classifier,  (c)(f) RF classifier. }}
\end{center}
\end{figure}

\subsection{\label{subsec:4:4} Source localization as a regression problem }

Source localization can be solved as a regression problem. For this problem, the output represents the continuous range and the input remains the vectorized covariance matrix Eq.~(\ref{SCM}). In this case, the highly nonlinear relationship between the covariance matrix elements and the range, caused primarily by modal interference, is difficult to fit a nonlinear model. In the training process, the input data remain the same, the labels are direct GPS ranges, and the weights and biases are trained using least-squares objective functions. 

The range predictions by FNN with different number of hidden layers along with the GPS ranges are given in Fig.~\ref{fig:FIG12}. Increasing the number of hidden layers increases the nonlinearity of the model and allows a larger number of parameters to be learned, thus significantly reducing the error for FNN regression. Figure~\ref{fig:FIG13} shows the results of SVM (Fig.~\ref{fig:FIG13}(a)(c)) and RF regressors (Fig.~\ref{fig:FIG13}(b)(d)) on two data sets. For these methods, since additional layers cannot be added to increase the nonlinearity, the performance lags FNN. The best MAPE values for each regressor are shown in Table~\ref{table:1}. Compared with classifiers, the FNN, SVM and RF degrade significantly for solving regression tasks.

\begin{figure}
\begin{center}
\includegraphics{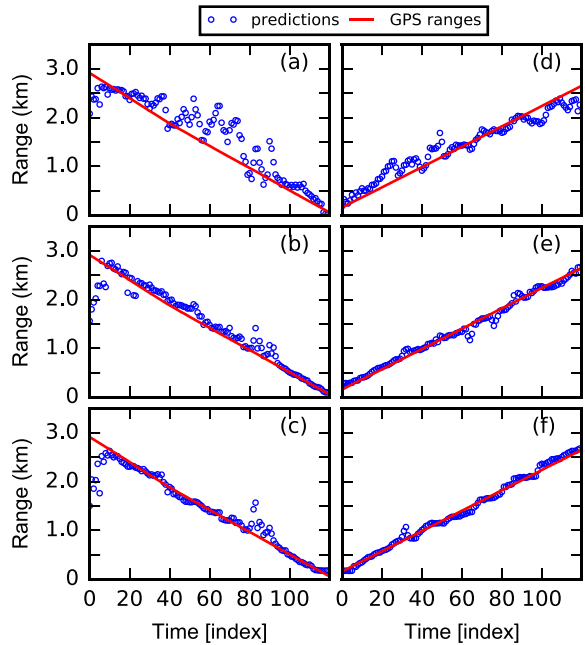}
\caption{ \label{fig:FIG12} {(Color online) Source localization as a regression problem. Range predictions on Test-Data-1 (a, b, c) and Test-Data-2 (d, e, f) by FNN for 300--950 Hz with 10 Hz increment, i.e. 66 frequencies. (a)(d) 1 hidden layer,  (b)(e) 2 hidden layers,  (c)(f) 3 hidden layers. Each hidden layer consists of 512 neurons.}}
\end{center}
\end{figure}

\begin{figure}
\begin{center}
\includegraphics{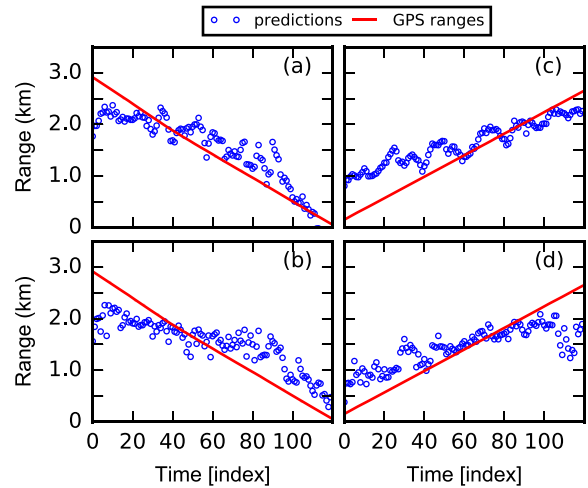}
\caption{ \label{fig:FIG13} {(Color online) Source localization as a regression problem. Range predictions on Test-Data-1 (a, b) and Test-Data-2 (c, d) by SVM and RF for 300--950 Hz with 10 Hz increment, i.e. 66 frequencies. (a)(c) SVM for regression,  (b)(d) RF for regression.}}
\end{center}
\end{figure}

\subsection{\label{subsec:4:5} Conventional matched-field processing }

The Bartlett MFP\cite{Jensen} is applied to Noise09 data for comparison. Two kinds of replica fields are used in the Bartlett processor. The first is generated by KRAKEN using the Noise09 environment in Fig. \ref{fig:FIG4}, with the corresponding ambiguity surfaces and maximum peaks shown in Fig. \ref{fig:FIG14}. We use the measured data (i.e. training data, 01/31/2009, 01:43--02:05) as the second group of replica fields as proposed in Ref. \onlinecite{Fialkowski}. The results are shown in Fig. \ref{fig:FIG15}. For each case, both single frequency (550 Hz) and broadband (300--950 Hz) are considered. 

From Figs. \ref{fig:FIG14} and \ref{fig:FIG15}, the Bartlett MFP fails to determine source positions using a single frequency, while the FNN still generates a number of reasonable predictions (see Fig. \ref{fig:FIG10}(a)). Despite improved performance using broadband MFP, there are some errors due to sidelobes. The MAPE of MFP predictions is shown in Table~\ref{table:1}. The minimum MAPE of Bartlett MFP is 19\% on Test-Data-1 and 30\% on Test-Data-2, which is much larger than the machine learning classifiers.
\begin{figure}
\begin{center}
\includegraphics{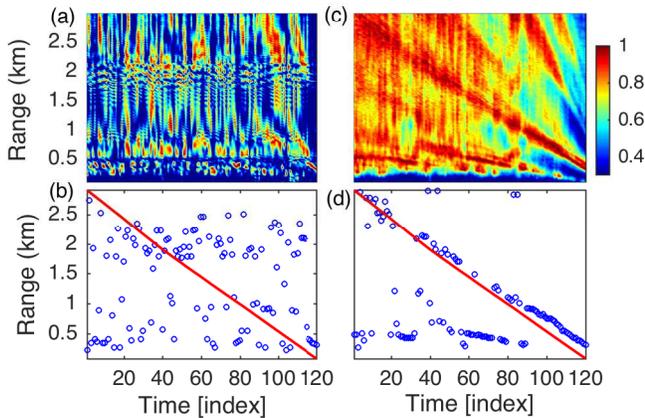}
\caption{ \label{fig:FIG14} (Color online) Localization using Bartlett matched-field processing based on synthetic replica fields on Test-Data-1. (a) ambiguity surface and  (b) maximum peaks for 550 Hz, (c) ambiguity surface and  (d) maximum peaks for 300--950 Hz with 10 Hz increment. Circles and solid lines denote predictions and GPS ranges respectively.}
\end{center}
\end{figure}

\begin{figure}
\begin{center}
\includegraphics{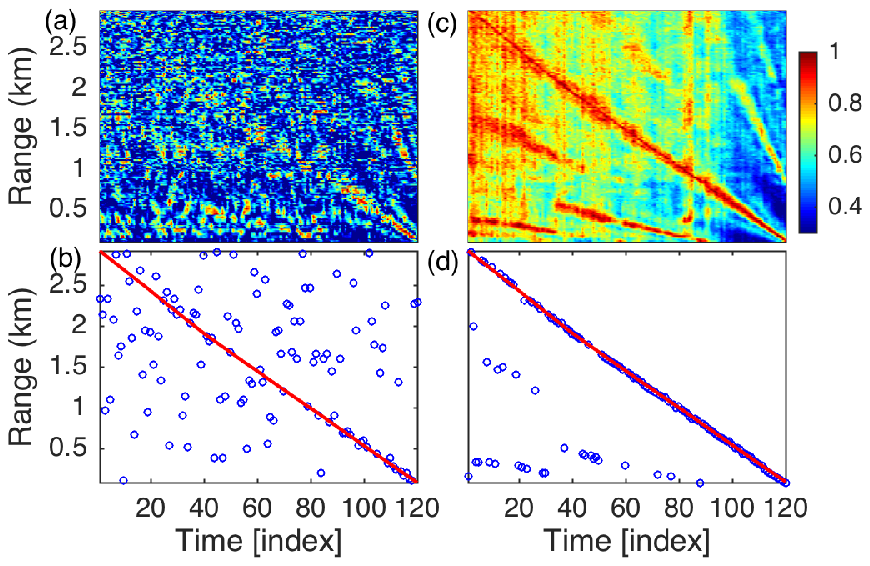}
\caption{ \label{fig:FIG15} (Color online) Localization using Bartlett matched-field processing (training data as replica fields) on Test-Data-1. (a) ambiguity surface and  (b) maximum peaks for 550 Hz, (c) ambiguity surface and  (d) maximum peaks for 300--950 Hz with 10 Hz increment. Circles and solid lines denote predictions and GPS ranges respectively.}
\end{center}
\end{figure}

\begin{table}[hbtp]
	\begin{center}
		\caption{\label{table:1}{Best MAPE rate of FNN, SVM, RF and MFP predictions.}}
		\begin{tabular}{  p{2.8cm}<{\centering}  p{2.5cm} <{\centering}  p{2.5cm} <{\centering}}
			\hline
			\hline
			\multirow{2}{*}{Model}
			& \multicolumn{2}{c}{~MAPE~ }\\
			&Test-Data-1 (\%) & Test-Data-2 (\%)\\
			\hline
			FNN classifier  & 3 &  3 \\
			%\hline
			SVM classifier &  2 & 2 \\
			%\hline
			RF classifier &  3 & 2 \\
			%\hline
			FNN regressor  & 10 &  5 \\
			%\hline
			SVM regressor &  42 & 59 \\
			%\hline
			RF regressor &  55 & 48 \\
			%\hline
			MFP  &  55 & 36 \\
			%\hline
			MFP with measured replica & \multirow{2}{*}{19} & \multirow{2}{*}{30} \\
			\hline
		\end{tabular}
	\end{center}
\end{table}

\section{\label{sec:5} Discussions}

\subsection{\label{subsec:5:1} Range resolution}

The number of classes, corresponding to the resolution of range steps, was varied to determine its effect on range estimation results. Previously (see Sec.~\ref{sec:4}) $K=138$ classes were used, corresponding to a range resolution of 20 m. The MAPE for predictions with different numbers of output classes by FNN, SVM and RF classifiers is given in Table~\ref{table:2} with 10 snapshots averaged for each sample. These three classifiers perform well for all tested range resolutions. 

\subsection{\label{subsec:5:2} Snapshots }

The number of snapshots averaged to create the SCMs may also affect the performance. Increasing the number of snapshots makes the input more robust to noise, but could introduce mismatch if the source is moving or the environment is evolving across the averaging period. The range estimation methods are tested using 1, 5, and 20 snapshots and the corresponding MAPE is shown in Table~\ref{table:2}. All of the three models degrade with 1 snapshot due to low SNR and become robust with more snapshots.

\subsection{\label{subsec:5:3} Number of hidden neurons and layers for FNN}

The MAPE of FNN with different numbers of hidden neurons and layers is given in Table~\ref{table:2}. Increasing the number of hidden neurons increases the number of parameters to fit in the FNN model. As a result, more of the variance in the data is captured. FNN has the minimum error when the number of hidden neurons is chosen as 128 or 2048 for Test-Data-1 (7\%) and 512 for Test-Data-2 (4\%). In the case of Test-Data-1, 256, 512 and 1024 have a similar result (8\%, 8\%, 8\%). As shown in Table~\ref{table:2}, the FNN with two hidden layers did not improve the prediction performance for classification.

\subsection{\label{subsec:5:4} Kernel and regularization parameters for SVM}

When using a Gaussian radial basis function kernel, the parameters $\gamma$ in Eq.~(\ref{RBF}) and the regularization parameter $C$ in Eq.~(\ref{svm_min2}) determine the best separation of the data by SVM. The MAPE versus these two parameters on two data sets is shown in Fig.~\ref{fig:FIG16}. As seen from the result, there exists an optimal interval for these two parameters (i.e. $10<C<10^3$ and $10^{-3}<\gamma<10^{-1}$). The SVM fails when $\gamma$ and $C$ are out of this interval, but is robust when $\gamma$ and $C$ are within the appropriate range.

\begin{figure}
	\begin{center}
		\includegraphics{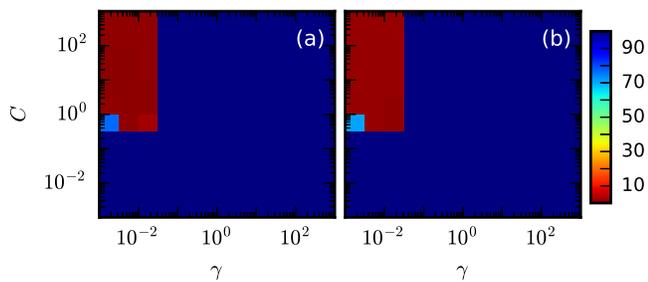}
		\caption{\label{fig:FIG16} (Color online) MAPE of the SVM classifier on (a) Test-Data-1 and (b) Test-Data-2. $\gamma$ is the kernel parameter and $C$ is the regularization parameter. There are 138 output classes and 10 snapshots averaged for each sample.}
	\end{center}
\end{figure}

\begin{figure}
	\begin{center}
		\includegraphics{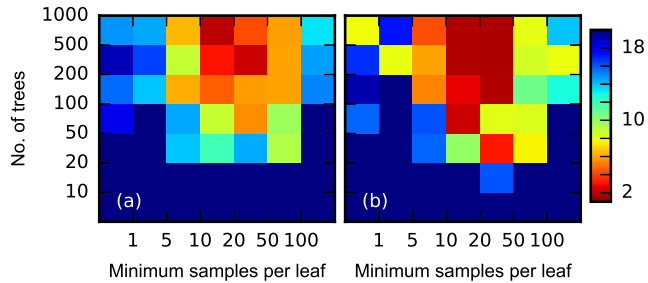}
		\caption{\label{fig:FIG17} (Color online) MAPE of the RF classifier versus the number of trees and the minimum samples per leaf on (a) Test-Data-1 and (b) Test-Data-2. There are 138 output classes and 10 snapshots averaged for each sample.}
	\end{center}
\end{figure}

\subsection{\label{subsec:5:5} Number of trees and minimum samples per leaf for RF}

The number of decision trees and the minimum samples per leaf\cite{Scikit} are the most sensitive parameters for the RF. Figure~\ref{fig:FIG17} shows the MAPE versus these two parameters. The RF parameters have a smaller range of possible values than SVM, but the RF classifier will not fail for any of these choices. The RF classifier has the best performance for more than 500 trees and 20 to 50 minimum samples per leaf.

\subsection{\label{subsec:5:6} CPU time}

The three machine learning models are efficient in CPU time (on a 2015--iMac). For the FNN with 1024 hidden neurons and 1000 training steps, it took 140 s for training and predicting the results. The SVM took 110 s, and the RF (500 trees and 50 minimum samples per leaf) was the fastest at 52 s.

\subsection{\label{subsec:5:7} Multiple sources and deep learning }

In our study, only one source is considered. The simultaneous multiple source localization problem is more challenging, especially for sources close to each other. Solving this problem with FNN is a multiple binary classification problem and will require additional training data.

Although the FNN with one hidden layer works well for the data sets in this paper, more complicated machine learning algorithms, e.g. deep learning, may be necessary for more complicated experimental geometries or ocean environments. 

\begin{table*}[hbtp]
\begin{center}
\caption{\label{table:2}{Parameter sensitivity of FNN, SVM and RF classifiers.}}
\begin{tabular}{ p{2.5cm}<{\centering} p{2.5cm} <{\centering}  p{2.5cm} <{\centering}    p{2.2cm} <{\centering}  p{2.5cm} <{\centering}  p{2.5cm} <{\centering}}
\hline
\hline
\multicolumn{6}{c}{ Part I: FNN classifier }    \\
\hline
{No. of hidden } & {No. of hidden } & {No. of  } & {No. of  }& \multicolumn{2}{c}{~MAPE~ }\\
 {layers} & {neurons} & {classes} & snapshots &Test-Data-1 (\%) & Test-Data-2 (\%)\\
 %\hline
%\hline
1 & 1024 & 1380 & 10  & 7 & 5\\
1 & 1024 & 690  & 10 & 3 & 6  \\
1 & 1024  & 276 & 10 & 6 & 8 \\
 1 & 1024 & 138 & 10  & 8 &  6 \\
1 & 1024  & 56 & 10 & 7 & 4 \\
1 & 1024  & 28 & 10 & 10 & 4 \\
1 & 1024  & 14 & 10 & 16 &  7\\
%\hline
1 & 1024 & 138 & 1  & 10 & 5 \\
1 & 1024 & 138  & 5 & 6 & 3 \\
1 & 1024  & 138 & 20 & 8 & 3 \\
%\hline
1 & 64  & 138 & 10 & 9 & 9 \\
1 & 128  & 138 & 10 & 7 & 7 \\
1 & 256  & 138 & 10 & 8 & 6 \\
1 & 512  & 138 & 10 &8 & 4 \\
1 & 2048  & 138 & 10 & 7 & 5 \\
%\hline
2 & 128& 138 & 10   & 9 & 8 \\
2 & 256 & 138 & 10  & 9 & 9 \\
2 & 512  & 138 & 10 & 6 & 8 \\
\hline
\multicolumn{6}{c}{ Part II: SVM classifier }    \\
\hline
 \multirow{2}{*}{$\gamma$}
 & \multirow{2}{*}{$C$} &{No. of  }  & {No. of  }&   \multicolumn{2}{c}{~MAPE~ }  \\
&  & classes & snapshots & Test-Data-1 (\%) & Test-Data-2 (\%) \\
%\hline
& & 1380& 10&2 &3 \\
& & 690&10&2 &3 \\
& & 276& 10&4 &3 \\
& & 138&10&2&2 \\
$10^{-2}$& 10 & 56&10&3 &3 \\
& & 28&10& 5&3 \\
& &138& 1&17& 5 \\
& &138& 5&2& 3\\
& &138& 20&3& 2\\
\hline
\multicolumn{6}{c}{ Part III: RF classifier }    \\
\hline
 \multirow{2}{*}{}
{No. of} & No. of samples &{No. of  }  & {No. of  }&   \multicolumn{2}{c}{~MAPE~ }  \\
trees & per leaf & classes & snapshots & Test-Data-1 (\%) & Test-Data-2 (\%) \\
%\hline
& & 1380& 10& 4 & 10 \\
& &690&10& 3 & 4 \\
& &276& 10& 3 & 3 \\
& & 138&10& 3 & 2\\
500 & 50 & 56&10& 9 & 5 \\
& & 28&10& 13 & 9\\
& &138& 1&20&15 \\
& &138& 5&6&5 \\
& &138& 20&3&2 \\
\hline
 \end{tabular}
\end{center}
\end{table*}

\section{\label{sec:6} Conclusion}

This paper presents an approach for source localization in ocean waveguides within a machine learning framework. The localization is posed as a supervised learning problem and solved by the feed-forward neural networks, support vector machines and random forests separately. Taking advantage of the modern machine learning library such as TensorFlow and Scikit-learn, the machine learning models are trained efficiently. Normalized sample covariance matrices are fed as input to the models. Simulations show that FNN achieves a good prediction performance for signals with SNR above 0 dB even with deficient training samples. Noise09 experimental data further demonstrates the validity of the machine learning algorithms.

Our results show that classification methods perform better than regression and MFP methods. The three classification methods tested (FNN, SVM, RF) all performed well with the best MAPE 2--3\%. The SVM performs well if the model parameters are chosen appropriately, otherwise it degrades significantly. By contrast, the FNN and RF are more robust on the choices of parameters.

The experimental results show that multi-frequency input generates more accurate predictions than single frequency (based on FNN). In the current study, the training and test data were from the same ship. In a realistic application, data from multiple ships of opportunity can be used as training data by taking advantage of the Automatic Identification System (AIS), a GPS system required on all cargo carriers. The tracks were quite similar and it would be interesting to compare the performance of the three methods as tracks deviate.

\begin{acknowledgments}
This work was supported by the Office of Naval Research under Grant No. N00014--1110439 and the China Scholarship Council.
\end{acknowledgments}

% --------------------------------------------------------------------------------------------------------------------------------

\vspace{0.2in}     %  included for better reading and an example of adding vertical space

 % -------------------------------------------------------------------------------------------------------------------
 %   Appendix  (optional)

\clearpage

%-------------------------------------------------------------------------------------------------------------
% List of Figure Captions
%-------------------------------------------------------------------------------------------------------------
\begin{itemize}

% Figure1 -  no bullets
 \item[]{Fig.~\ref{fig:FIG1}. (a) Diagram of a feed-forward neural network and (b) Sigmoid function.}

 \item[]{Fig.~\ref{fig:FIG2}. (Color online) A linear hyperplane learned by training an SVM in two dimensions ($D=2$).}

 \item[]{Fig.~\ref{fig:FIG3}. (Color online) Decision tree classifier and corresponding rectangular regions shown for two--dimensional data with $K=2$ classes ($D=2$, $M=3$) and 1000 training points.}

 \item[]{Fig.~\ref{fig:FIG4}. (Color online) (a) Waveguide parameters and source-receiver configuration. (b) Sound speed profile of water column.}

 \item[]{Fig.~\ref{fig:FIG5}. (Color online) Range predictions by FNN on test data set with SNR of (a) $-$10, (b) $-$5, (c) 0, and (d) 5 dB.}

 \item[]{Fig.~\ref{fig:FIG6}. (Color online) Output probability for range 0.1--2.86 km (the true range is 1.5 km) after training steps (1, 100, 200, 300). The top line represents the target output.}

 \item[]{Fig.~\ref{fig:FIG7}. (Color online) Cross entropy Eq.~(\ref{cross_entropy}) versus optimization steps on training (solid) and test (dashed) data with SNR of (a) $-$10, (b) $-$5, (c) 0, and (d) 5 dB.}

 \item[]{Fig.~\ref{fig:FIG8}. (Color online) Ship tracks for Noise09 experiment during the periods (a) 01/31/2009, 01:43--02:05 (training data, ship speed 2 m/s), (b) 01/31/2009, 01:05--01:24 (Test-Data-1, ship speed $-$2 m/s), and (c) 02/04/2009, 13:41--13:51 (Test-Data-2, ship speed 4 m/s). }

 \item[]{Fig.~\ref{fig:FIG9}. (Color online) Spectra of shipping noise during periods (a) 01/31/2009, 01:43--02:05, (b) 01/31/2009, 01:05--01:24, and (c) 02/04/2009, 13:41--13:51. }

 \item[]{Fig.~\ref{fig:FIG10}. (Color online) Range predictions on Test-Data-1 (a, b, c) and Test-Data-2 (d, e, f) by FNN. (a)(d) 550 Hz,  (b)(e) 950 Hz,  (c)(f) 300--950 Hz with 10 Hz increment, i.e. 66 frequencies. The time index increment  is 10 s for Test-Data-1, and 5 s for Test-Data-2.}

 \item[]{Fig.~\ref{fig:FIG11}. (Color online) Source localization as a classification problem. Range predictions on Test-Data-1 (a, b, c) and Test-Data-2 (d, e, f) by FNN, SVM and RF for 300--950 Hz with 10 Hz increment, i.e. 66 frequencies. (a)(d) FNN classifier, (b)(e) SVM classifier,  (c)(f) RF classifier.}

\item[]{Fig.~\ref{fig:FIG12}. (Color online) Source localization as a regression problem. Range predictions on Test-Data-1 (a, b, c) and Test-Data-2 (d, e, f) by FNN for 300--950 Hz with 10 Hz increment, i.e. 66 frequencies. (a)(d) 1 hidden layer,  (b)(e) 2 hidden layers,  (c)(f) 3 hidden layers. Each hidden layer consists of 512 neurons.}

 \item[]{Fig.~\ref{fig:FIG13}. (Color online) Source localization as a regression problem. Range predictions on Test-Data-1 (a, b) and Test-Data-2 (c, d) by SVM and RF for 300--950 Hz with 10 Hz increment, i.e. 66 frequencies. (a)(c) SVM for regression,  (b)(d) RF for regression.}

 \item[]{Fig.~\ref{fig:FIG14}. (Color online) Localization using Bartlett matched-field processing based on synthetic replica fields. (a) ambiguity surface and  (b) maximum peaks for 550 Hz, (c) ambiguity surface and  (d) maximum peaks for 300--950 Hz with 10 Hz increment. Circles and solid lines denote predictions and GPS ranges respectively.}
 
 \item[]{Fig.~\ref{fig:FIG15}. (Color online) Localization using Bartlett matched-field processing (training data as replica fields). (a) ambiguity surface and  (b) maximum peaks for 550 Hz, (c) ambiguity surface and  (d) maximum peaks for 300--950 Hz with 10 Hz increment.  Circles and solid lines denote predictions and GPS ranges respectively.}
   
\item[]{Fig.~\ref{fig:FIG16}. (Color online) MAPE of the SVM classifier on (a) Test-Data-1 and (b) Test-Data-2. $\gamma$ is the kernel parameter and $C$ is the regularization parameter. There are 138 output classes and 10 snapshots averaged for each sample.}
   
\item[]{Fig.~\ref{fig:FIG17}. (Color online) MAPE of the RF classifier versus the number of trees and the minimum samples per leaf on (a) Test-Data-1 and (b) Test-Data-2. There are 138 output classes and 10 snapshots averaged for each sample.}

\end{itemize}

\end{space}

\end{document}